\documentclass[twocolumn,showpacs,preprintnumbers,amsmath,amssymb,superscriptaddress]{revtex4}

\usepackage{calc}
\usepackage{float}
\usepackage{graphicx}% Include figure files
\usepackage{bm}% bold math
\def \equi#1{\mathrel{\mathop{\kern 0pt\sim}\limits_{#1}}} 
\def \equiqui#1{\mathrel{\mathop{\kern 0pt\simeq}\limits_{#1}}} 
\scrollmode
\begin{document}

\title{Active transport in  dense diffusive single-file systems}

\author{P. Illien}
\affiliation{Laboratoire de Physique Th\'eorique de la Mati\`ere Condens\'ee, CNRS UMR 7600, case courrier 121, Universit\'e Paris 6, 4 Place Jussieu, 75255
Paris Cedex France} 

\author{O. B\'enichou}
\affiliation{Laboratoire de Physique Th\'eorique de la Mati\`ere Condens\'ee, CNRS UMR 7600, case courrier 121, Universit\'e Paris 6, 4 Place Jussieu, 75255
Paris Cedex France}

\author{C. Mej\'{\i}a-Monasterio}
\affiliation{Laboratory of Physical Properties,
Technical University of Madrid, Av. Complutense s/n, 28040 Madrid, Spain, and
Department of Mathematics and Statistics,
    University of  Helsinki, P.O.  Box 68  FIN-00014 Helsinki, Finland}

\author{G. Oshanin}
\affiliation{Laboratoire de Physique Th\'eorique de la Mati\`ere Condens\'ee, CNRS UMR 7600, case courrier 121, Universit\'e Paris 6, 4 Place Jussieu, 75255
Paris Cedex France}

\author{R. Voituriez}
\affiliation{Laboratoire de Physique Th\'eorique de la Mati\`ere Condens\'ee, CNRS UMR 7600, case courrier 121, Universit\'e Paris 6, 4 Place Jussieu, 75255
Paris Cedex France}

\date{\today}

\begin{abstract}

We study a minimal model of active transport in crowded single-file environments which generalises the emblematic model of single file diffusion to the case when the tracer particle (TP) performs either an autonomous directed motion or is biased by an external force, while all other particles of the environment (bath) perform unbiased diffusions.  We 
derive explicit expressions, valid   in the limit of high density of bath particles,  of the full 
 distribution $P_n(X)$ of the TP position and of all its cumulants,  
 for arbitrary values of the bias $f$ and for any time $n$. Our analysis reveals striking features, such as the anomalous scaling $\propto\sqrt{n}$ of all cumulants, the equality of cumulants of the same parity characteristic of a Skellam distribution and a convergence to a Gaussian distribution in spite of  asymmetric density profiles of bath particles. 
 %We believe that these characteristics  provide a signature
 %of active transport in dense diffusing single file systems and could make possible its experimental identification and quantitative analysis.
 Altogether, our results provide the full statistics of the TP position, and set the basis 
  for a refined analysis of real trajectories of active particles in  crowded single-file environments. 
 \end{abstract}


\maketitle

{\it Introduction.} Single-file diffusion refers to  one-dimensional diffusion of interacting 
particles  that  can not by-pass each other. Clearly, in such a geometry 
the initial order of particles remains the same over time, and 
this very circumstance appears so crucial that the movements of individual particles become strongly 
correlated:  the displacement of any given tracer  particle (TP) on progressively larger distances
 necessitates the motion of more and more other particles in the same direction. 
 This results in 
 a subdiffusive growth 
 of the TP
 mean-square displacement $\overline{X^2} \sim \sqrt{t}$, first discovered analytically by Harris \cite{Harris:1965} and subsequently re-established for systems with differently organised dynamics (see, e.g., Refs.\cite{Levitt:1973,Fedders:1978,Alexander:1978,Arratia:1983,Lizana:2010,Taloni:2008,Gradenigo:2012}).
 Nowadays, a single-file diffusion, prevalent in many physical, chemical and biological processes and experimentally evidenced by passive microrheology   in 
zeolites, transport of confined colloidal particles or charged spheres in circular channels \cite{Gupta:1995,Hahn:1996,Wei:2000,Meersmann:2000,Lin:2005}. It  provides a paradigmatic example 
 of anomalous diffusion in crowded \textit{equilibrium} systems, 
 which emerges due to a cooperative many-particle behavior.

On the other hand,   systems that consume energy for  propulsion -- {\it active} particle systems -- have received growing attention in the last decade, both because of the new physical phenomena that they display and their wide range of applications. Examples include self-propelled particles such as molecular motors or motile living cells 
 \cite{Toner:2005}, and externally driven particles, such as  
probes in active microrheology experiments \cite{Wilson:2011a}. The intrinsic out-of-equilibrium
nature of these systems leads to remarkable effects such as non-Boltzmann
distributions \cite{Puglisi:1999}, long-range order even in low spatial
dimensions   \cite{Toner:1995} and spontaneous flows
\cite{Voituriez:2005}. In particular, 1D assemblies of active particles have been extensively studied in the context of TASEP models.

\begin{figure}[ht!]
\begin{center}
 \includegraphics[width=8cm]{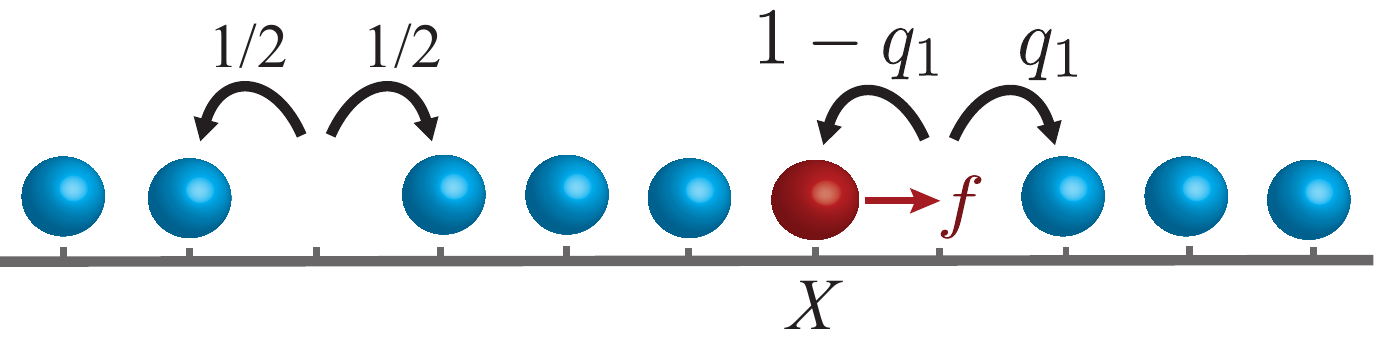}
\end{center}
\caption{(color online) Model notations. }\label{fig0}
\end{figure}

However, up to now, active transport in  diffusive single file systems,   which involve an active TP performing an autonomous directed motion or pulled by a constant external force $f$ in a 
1D bath of  unbiased 
diffusive particles with hard-core interactions,  has drawn uncomparably less attention \footnote{Note that  this model is very different from   TASEP inspired systems where {\it all} particles are  biased.}. 
Such dynamics, depicted in figure \ref{fig0}  provides a minimal model of active transport in  crowded single-file environments,
which schematically mimics situations as varied as  the active transport of a vesicle in a crowded axone \cite{Loverdo:2008},  directed cellular movements in crowded channels \cite{Hawkins:2009a} or active-microrheology in capillaries \cite{Wilson:2011a}.   
In this context, the only available theoretical results concern 
the large time behavior of the
 mean displacement $\overline{X}$ of the  TP, which has been shown to grow sub-linearly with time  $\overline{X} \sim  \sqrt{t}$ \cite{Burlatsky:1992a,Burlatsky:1996b,Landim:1998a,BookEinstein}.
In fact,
the biased 
TP drives the bath particles  to a \textit{ non-equilibrium} 
state with an asymmetric distribution:  
the bath particles accumulate in front of the TP thus increasing the frictional force, 
and 
are depleted behind. The  extent of these perturbations   grows in time in proportion to  $\sqrt{t}$ and characterizes a subtle 
 interplay between the bias, 
formation of non-equilibrium density profiles and backflow effects of the medium on the TP. 
 
In this Letter  we focus on this minimal model of active transport in  diffusive single file systems.
 Going beyond the previous 
 analysis of the TP mean displacement, we present  in the limit of high density of bath particles exact expressions of the full 
 distribution $P_n(X)$ of the TP position and of all its cumulants,  
 for arbitrary values of the bias $f$ and for any time $n$. 
 %In addition, the detailed analysis of the cumulants in this large density limit
 %yields the following results :
%(i)   all even 
 %cumulants are equal to each other for any moment of time $n$, and so do all odd cumulants;
%(ii) 
% all cumulants grow in proportion to $\sqrt{n}$ in the leading order in $n$, and,  
 %this leading large-$n$ behaviour of
 %the cumulants  is \textit{independent} 
 %of $f$ when $j$ is even, and does depend on $f$ for odd $j$.  
In particular,  in this high density limit,  it is shown that at large times this distribution is  Gaussian, with
 mean  $\overline{X} \sim \alpha_f(\rho) \sqrt{n}$ and  variance 
  growing asymptotically as $\sqrt{n}$. Remarkably, in this limit the variance is  proved to be independent of $f$. Altogether, our results provide the full statistics of the TP position, and set the basis 
  for a refined analysis of real trajectories of active particles in  crowded single-file environments. 

{\it The model}. Consider a one-dimensional, infinite in both directions line of integers $x$, 
populated by hard-core particles present at mean density $\rho$, performing symmetric random walks.    At $t = 0$ we introduce at the origin of the lattice an active  TP, hopping on its right (resp. left) neighbor site with probability  $p_1$ (resp. $p_{-1}$), which process is also
constrained by  hard-core exclusion. In what follows, we focus on the limit of a dense system, corresponding to the limit of a small vacancies density  $\rho_0 =1-\rho \ll 1$. In this limit it is most convenient to 
follow the vacancies, rather than the particles. We thus formulate directly the  dynamics of the vacancies, which unambiguously defines the full dynamics of the system.  Following \cite{Brummelhuis:1989a,Benichou:2002qq}, we assume that at each time step each  vacancy is moved  to one of  its nearest neighbours sites, with equal probability. As long as a vacancy  is surrounded only by bath particles, it thus performs a symmetrical nearest neighbors  random walk. However, due to the biased nature of the movement of the TP, specific rules have to be defined when a vacancy is adjacent to the TP. In this case, if the vacancy  occupies the site to the right (resp. to the left) of the TP, 
we stipulate that it has a probability $q_1=1/(2p_1+1)$ (resp. $q_{-1}=1/(2p_{-1}+1)$) to jump to the right (resp. to the left) and $1-q_1$ (resp. $1-q_{-1}$) to jump to the left (resp. to the right). These rules are the discrete counterpart of a continuous time version of the model \footnote{In the continuous time model,  waiting times of particles are exponentials with mean 1. In that case, $q_1$ is in fact the probability that the adjacent  bath particle jumps onto the vacancy before the TP}, as shown in  \cite{Benichou:2002qq}.
Note that  a complete description of the dynamics would requires  additional rules for cases where two vacancies are adjacent or have common neighbours; however, these cases contribute only to ${\mathcal O}(\rho^2)$, and can thus be left  unstated.

{\it Single file  with a single vacancy}. 
We start with an auxiliary problem 
in which the system contains just a single vacancy initially at position $Z$, and which will be proved next to be a key step in the resolution of the complete problem with a (small) concentration of vacancies. 
Let $p^n_{ Z}(X)$
denote the probability of having 
the TP at site $X$ at time moment $n$, 
given that the vacancy commenced its random walk at $Z$. Clearly, in a single vacancy case
 this probability is not equal to zero only for $X = 0$ and $X = 1$, if $Z > 0$, and $X = 0$ and $X = - 1$, if $Z < 0$. 
Following Refs. \cite{Brummelhuis:1989a,Benichou:2002qq}, we then  represent 
$p^n_{ Z}(X)$ as:
\begin{eqnarray}
&&p_{Z}^n(X)=\delta_{ X,{\bf 0}}\left(1-\sum_{j=0}^n F_Z^j\right) + \nonumber\\
&+&\sum_{p=1}^{+\infty}\sum_{m_1,m_2, \ldots,m_p=1}^{+\infty}
\sum_{m_{p+1}=0}^{+\infty}\delta_{m_1+\ldots+m_{p+1},n}\delta_{X,\frac{{\rm sign}(Z)+(-1)^{p+1}}{2}}  \nonumber\\
&\times& \left(1-\sum_{j=0}^{m_{p+1}}F_{(-1)^p}^j\right) \times F_{(-1)^{p+1}}^{m_p}\ldots F_{-1}^{m_2} F_Z^{m_1} ,
\label{Ptr}
\end{eqnarray}
where  $\delta_{a,b} = 1$ when $a = b$ and is equal to zero, otherwise,
and $F^n_{Z}$ is the probability that the 
vacancy, which started its random walk at site Z, 
arrived to the origin for the first time at time moment $n$.

Let now $g(\xi)$ denote the generating function of any time-dependent function $g(n)$, 
$g(\xi)\equiv\sum_{n=0}^\infty g(n) \xi^n$. Then, 
Eq. (\ref{Ptr}) implies that the generating function of the 
propagator of the single-vacancy model can be expressed via the generating functions of the corresponding first-passage distributions as   
\begin{equation}
p_{\pm1}(X;\xi)=\frac{\delta_{X,0}(1-F_{\pm1})+\delta_{X,\pm1}F_{\pm1}(1-F_{\mp1})}{(1-F_1F_{-1})(1-\xi)},
\end{equation}
where we have used the short notations $F_{\pm1}\equiv F_{\pm1}(\xi)$.

{\it Single file  with a small concentration of vacancies}. 
We now turn to the original problem with a small but finite
 density $\rho_0$ of vacancies and aim to express 
 the desired probability $P_n(X)$ of finding the TP at site $X$ at time $n$ via the propagator for a single-vacancy problem. 
 We consider first a finite chain with $L$ sites, $M$ of which 
 are vacant and the initial positions of the latter are denoted by $Z_j$, $j=1, \ldots,M$. 
Then, the probability $P_n(X|\{Z_{j}\})$ of finding the TP 
at position $X$ at
time moment $n$ as a result
of its interaction with all the vacancies collectively, for their fixed initial configuration, writes
 \begin{equation}
P_n(X|\{Z_{j}\}) = \sum_{Y_1,Y_2, \ldots,Y_M} \delta_{X, Y_1+\ldots+Y_M} P_n(\{Y_j\}|\{Z_{j}\}),
 \end{equation}
  where $P_n(\{Y_j\}|\{Z_{j}\})$ stands for the conditional probability
 that within the time interval $n$ the TP has performed a 
 displacement $Y_1$ due to interactions with the first vacancy, a displacement
 $Y_2$ due to the interactions with the second vacancy, and etc.  In the
 lowest order in the density of vacancies, the vacancies contribute
 independently to the total  displacement of the tracer, so that the latter conditional probability decomposes
  \begin{equation}
P_n(\{Y_j\}|\{Z_{j}\})\equi{\rho_0\to0}\prod_{j=1}^M  p^n_{Z_j}(Y_j) \,,
 \end{equation}
   where
$p^n_{Z_j}(Y_j)$ is the single-vacancy propagator and the symbol 
$\equi{\rho_0\to0}$ signifies the leading behavior in the small density of vacancies limit.
Note that such an approximation  yields results which are exact 
to the order $\mathcal{O}(\rho_0)$, and hence, such a description
 is expected to be
quite accurate when $\rho_0 \ll 1$ \cite{Brummelhuis:1989a,Benichou:2002qq}.
Next, we suppose that initially the vacancies are uniformly distributed 
on the chain (except for the origin, which is occupied by the TP) and
average $P_n(X|\{Z_{j}\})$ over the initial distribution of the vacancies. In doing so and
subsequently turning
to the thermodynamic limit, i.e. setting 
$ L\to\infty$, $M\to\infty$ with $M/L=\rho_0$ kept fixed, we find
 that the generating function of the second characteristic function 
\begin{equation}
\psi_X(k;\xi)\equiv \sum_{n=0}^\infty \ln(\widetilde{P}^n(k)) \xi^n
\end{equation}
satisfies
\begin{equation}
\label{psi}
\lim_{\rho_0\to0} \frac{\psi_X(k;\xi)}{\rho_0} = -\sum_{\epsilon=\pm1}\left(\frac{1}{1-\xi}-\widetilde{p}_{-\epsilon}(k;\xi)e^{i\epsilon k }\right)\sum_{Z=1}^\infty F_{\epsilon Z}(\xi),
\end{equation}
Our last step consists in the explicit determination of $F_{\pm1}$ and $\sum_{Z=1}^\infty F_{\epsilon Z}(\xi)$ in Eq.(\ref{psi}). We note that both can be readily expressed via 
the first-passage time density at the origin at time $n$ of a symmetric one dimensional Polya random walk, starting at time 0 at position $l$, denoted as  $f^n_l$, since, by 
partitioning over the first time when 
the sites adjacent to the origin are reached, we have:
\begin{equation}
\label{z}
F_{\pm1}^n=(1-q_{\pm1})\delta_{n,1}+q_{\pm1}\sum_{k=1}^n f^{k-1}(1) F_{\pm1}^{n-k}.
\end{equation}
Multiplying both sides of Eq. (\ref{z}) by $\xi^n$, performing summation over $n$ and taking into account that  $f_l(\xi) = \sum f_l^n  \xi^n = ((1 - \sqrt{1-\xi^2})/\xi)^{|l|}$ \cite{Hughes:1995}, we find that 
\begin{equation}
F_{\pm1}=\frac{(1-q_{\pm1})\xi}{1-q_{\pm1}(1-\sqrt{1-\xi^2})}.
\end{equation}
Similarly, noticing that
\begin{equation}
F^{n}_{Z}=\begin{cases}&\sum_{k=1}^n f_{Z-1}^kF^{n-k}_{1} \;{\rm if}\; Z>0,\\
&\sum_{k=1}^n f_{-1-Z}^kF^{n-k}_{-1} \;{\rm if}\; Z<0,
\end{cases}
\end{equation}
and using the definition of $f_l(\xi)$ given above, we obtain 
\begin{equation}
\label{zz}
\sum_{Z=1}^\infty F_{\pm Z}(\xi)=\frac{F_{\pm1}}{1-(1-\sqrt{1-\xi^2})/\xi}.
\end{equation}
Gathering the results in Eqs. (\ref{z}) to (\ref{zz}), substituting them into Eq.(\ref{psi}), we 
finally derive our central analytical result which defines the exact  (in the leading in $\rho_0$ order) 
generating function of the cumulants of arbitrary order $j$:
\begin{equation}
\label{gen2}
\lim_{\rho_0\to0} \frac{\kappa_{(j)}(\xi)}{\rho_0} = \frac{F_1(1-F_{-1})+(-1)^jF_{-1}(1-F_1)}{(1-\xi)(1-(1-\sqrt{1-\xi^2})/\xi)(1-F_1F_{-1})} \,.
\end{equation}
This result gives access to the full statistics of the position of the TP and puts forward striking characteristics of active transport in dense diffusive single file systems as detailed below.
   
(i) First conclusion we can draw from Eq. (\ref{gen2}) is that for arbitrary $f$ (including $f =0$) all odd cumulants have the same generating function $\kappa_{\rm odd}(\xi)$, and all even cumulants have the same generating function $\kappa_{\rm even}(\xi)$. This means that at any moment of time and for any $f$ 
all cumulants $\kappa_{(j)}(n)$ with arbitrary odd $j$ are equal to each other, $\kappa_{(2 j +1)}(n) = \kappa_{\rm odd}(n)$, and so do all the cumulants with arbitrary even $j$, $\kappa_{(2 j)}(n) = \kappa_{\rm even}(n)$.

Parenthetically, we note that, in the classical case of single file diffusion (i.e. $f =0$), the generating function in Eq.(\ref{gen2}) can be inverted explicitly to give $\kappa_{\rm odd}(n) \equiv 0$ and for arbitrary time moment $n$
\begin{equation}
\label{1d}
\lim_{\rho_0\to0} \frac{\kappa_{\rm even}(n)} {\rho_0} =
\frac{2}{\sqrt{\pi}} \, \frac{\Gamma\left(\lfloor{\frac{n-1}{2}}\rfloor + \frac{3}{2}\right)}{\Gamma\left(\lfloor{\frac{n-1}{2}}\rfloor +1\right)} \,,
\end{equation}
where $\Gamma(\cdot)$ is the Gamma function and $\lfloor{x}\rfloor$ is the floor function. This expression, which can be shown to be compatible with the well-known Gaussian form in the large time limit, seems to be new.

(ii) Second, turning to the limit $\xi \to 1$ (large-$n$ limit) we find the leading in time asymptotic behavior of the cumulants of arbitrary order: 
\begin{equation}
\label{g}
\lim_{\rho_0\to0} \frac{\kappa_{(2j+1)}^{n}}{\rho_0}= (p_1-p_{-1})  \sqrt{\frac{2n}{\pi}}-2p_1p_{-1}(p_1-p_{-1})+o(1)
\end{equation}
\begin{equation}
\label{gg}
\lim_{\rho_0\to0} \frac{\kappa_{(2j)}^{n}}{\rho_0} = \sqrt{\frac{2n}{\pi}}+o(1) \,, \,\,\, j=0,1,2, \ldots .
\end{equation}
Equations (\ref{g}) and (\ref{gg}) signify that, remarkably, the leading in time behavior of all even cumulants is \textit{independent} of the force $f$, 
while the leading in time 
behavior of all odd cumulants does depend on $f$. In addition, for the standard choice of the transition 
probabilities such that $p_1 = 1- p_{-1}$ and
$p_1/p_{-1} = \exp(\beta f)$, where $\beta$ is the reciprocal temperature, and for the specific case $j=0$, 
we check from Eq. (\ref{g}) that
%\begin{equation}
%\label{ggg}
%\frac{\kappa_{(2j+1)}(n)}{\rho_0}  \equi{\rho_0\to0}  \tanh\left(\beta f/2\right)  \sqrt{\frac{2n}{\pi}}  -
%\frac{ \tanh\left(\beta f/2\right) }{2 \cosh\left(\beta f/2\right) }+o(1)\,,
%\end{equation}
\begin{equation}
\lim_{\rho_0\to0} \frac{{\overline X}}{\rho_0} =\tanh(\beta f/2)\sqrt{2n/\pi},
\end{equation}
which reproduces, for $j =0$, the results of  \cite{Burlatsky:1996b} and \cite{Landim:1998a}. Note that this anomalous scaling $\propto \sqrt{n}$ holds for all cumulants.

(iii) We finally provide an explicit expression of the full   distribution function $P_n(X)$ for any $n$.
As a matter of fact,  the equality at leading order in $\rho_0$ of cumulants of the same parity proved in point (i) shows that  the distribution associated to these cumulants is  of  Skellam type \cite{skellam}, so that :\begin{eqnarray}
\label{diss}
P_n(X) &\equiqui{\rho_0\to0}& \exp\left(-\kappa_{\rm even}(n)\right) \left(\frac{\kappa_{\rm even}(n) + \kappa_{\rm odd}(n)}{\kappa_{\rm even}(n) - \kappa_{\rm odd}(n)}\right)^{X/2} \, \nonumber\\
&\times& I_X\left(\sqrt{\kappa_{\rm even}^2(n) - \kappa_{\rm odd}^2(n)}\right) \,,
\end{eqnarray}
where $I_X(\cdot)$ is the modified Bessel function.  Importantly, we find that despite the known asymmetry of the concentration profile of the bath particles \cite{Burlatsky:1996b}, the rescaled variable
$(X_n-\kappa_{\rm odd}(n))/\sqrt{\kappa_{\rm even}(n)}$ is asymptotically distributed accordingly to  a normal law.  More precisely, 
the convergence to this Gaussian distribution can be quantified by the skewness $\gamma_1 = \kappa_{(3)}(n)/\kappa_{(2)}^{3/2}(n)$ and excess kurtosis $\gamma_2 = \kappa_{(4)}(n)/\kappa_{(2)}^{2}(n)$ 
 of the distribution $P_n(X)$. From Eqs.(\ref{g}) and (\ref{gg}) we readily find 
that in the leading in $\rho_0$ order 
\begin{equation}
\gamma_1  \equi{\rho_0\to0}  \frac{\tanh\left(\beta f/2\right)}{\rho^{1/2}_0} \, \left(\frac{\pi}{2 n}\right)^{1/4} + o(1/n^{1/4}) \,,
\end{equation}
and 
\begin{equation}
\gamma_2  \equi{\rho_0\to0}  \frac{1}{\rho_0} \, \left(\frac{\pi}{2 n}\right)^{1/2} + o(1/n^{1/2}) \,.
\end{equation}
Note that $\gamma_1 > 0$ which signifies that 
the right tail of $P_n(X)$ 
is longer and 
the fluctuations are more pronounced for $X > \overline{X}$ where the bath particles accumulate, than in the region $X < \overline{X}$ depleted with the bath particles.

\begin{figure}[ht!]
\begin{center}
 \includegraphics[width=8cm]{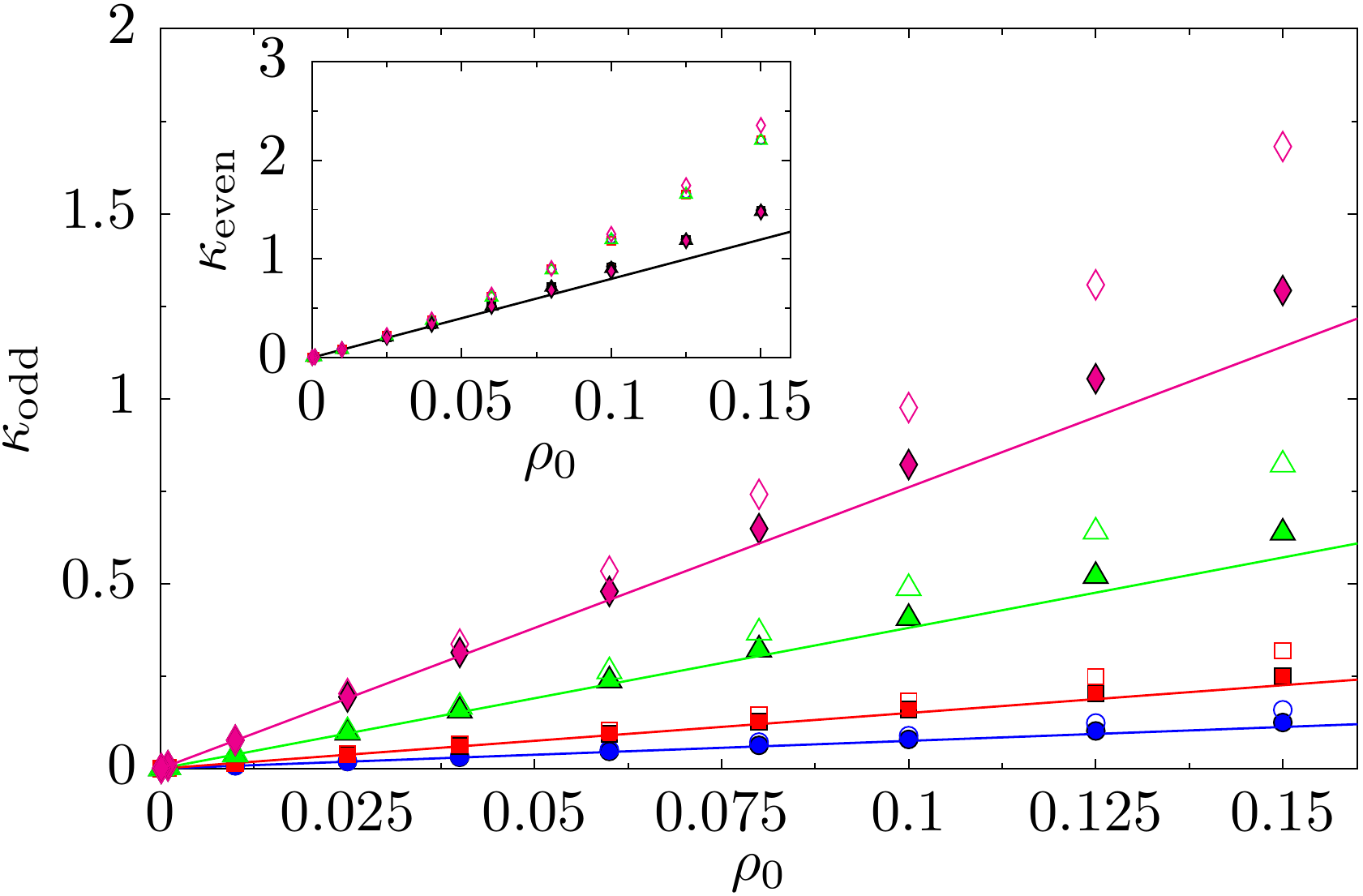}
\end{center}
\caption{(color online) 
%a) $Discrete- vs continuous-time dynamics
%in the passive case ($f=0$) for $\rho_0 = 0.002$. Red curve represents our   
%Eq. (\ref{1d}). Symbols are the results of numerical simulations with
%discrete- (circles) and continuous-time (crosses) dynamics. 
 Odd cumulants at time $n = 100$ vs 
$\rho_0$. The straight lines define our predictions in 
Eq. (\ref{gen2}) for different values of $p_1$, while the filled and empty symbols 
  are the results of numerical simulations for the first and third cumulants, respectively. Circles are results for $p_1 = 0.55$, squares - for $p_1 = 0.6$, triangles - for $p_1 = 0.75$ and diamonds - for $p_1 = 0.98$. The inset shows analogous results for the second and the fourth cumulants.}\label{fig1}
\end{figure}

\begin{figure}[h!]
\begin{center}
      \includegraphics[width=9cm]{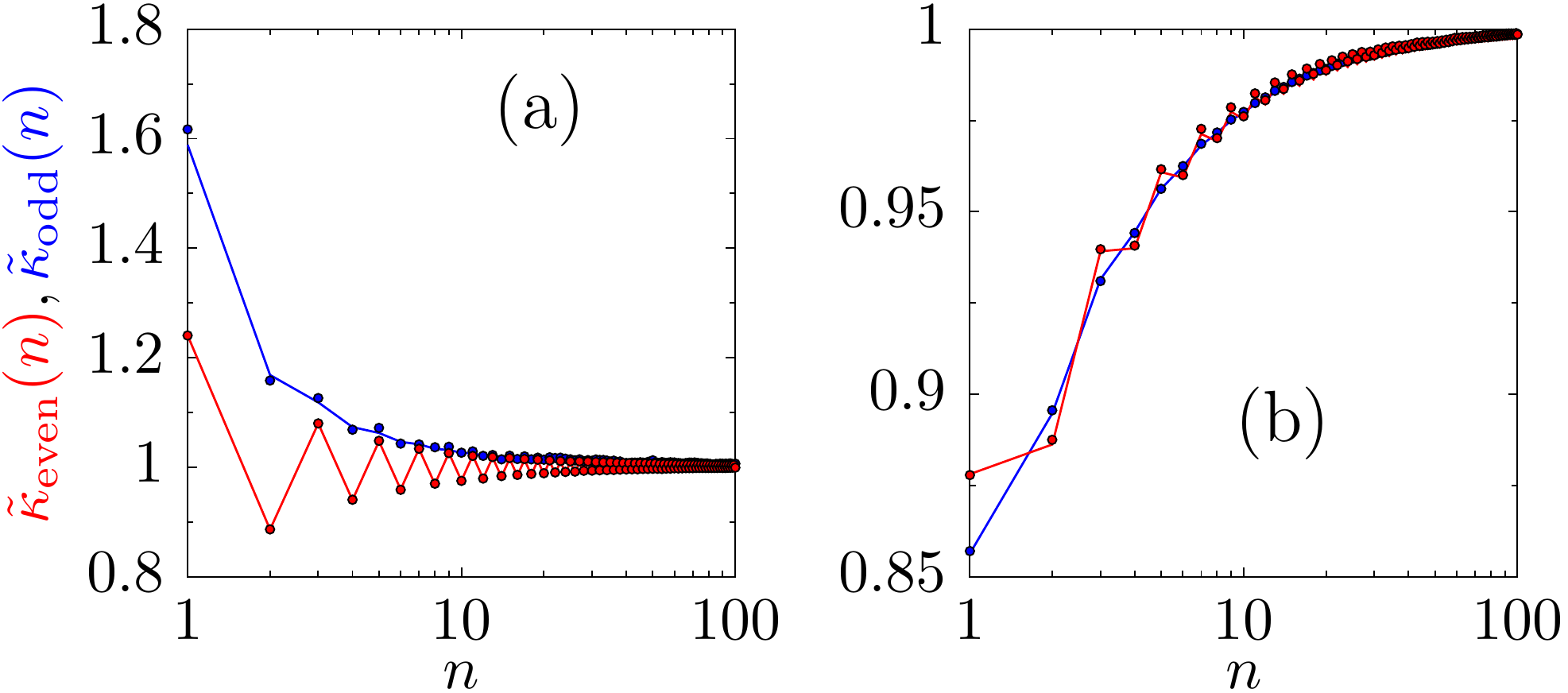}
\end{center}
\caption{(color online) Reduced cumulants $\tilde{\kappa}_{(\rm even)}(n) =  \kappa_{(\rm even)}(n)/ \sqrt{2 n/\pi}$ 
and $\tilde{\kappa}_{(\rm odd)}(n) =  \kappa_{(\rm odd)}(n)/[ (p_1 - p_{-1}) \sqrt{2 n/\pi}-2p_1p_{-1}(p_1-p_{-1})]$ vs time $n$ for $\rho_0 = 0.01$ and a) $p_1 = 0.6$ and b) $p_1 = 0.98$. Solid lines give the results of the inversion of  Eq. (\ref{gen2}), while symbols are the results of numerical simulations.}\label{fig2}
\end{figure}

\begin{figure}[h!]
\begin{center}
      \includegraphics[width=8cm]{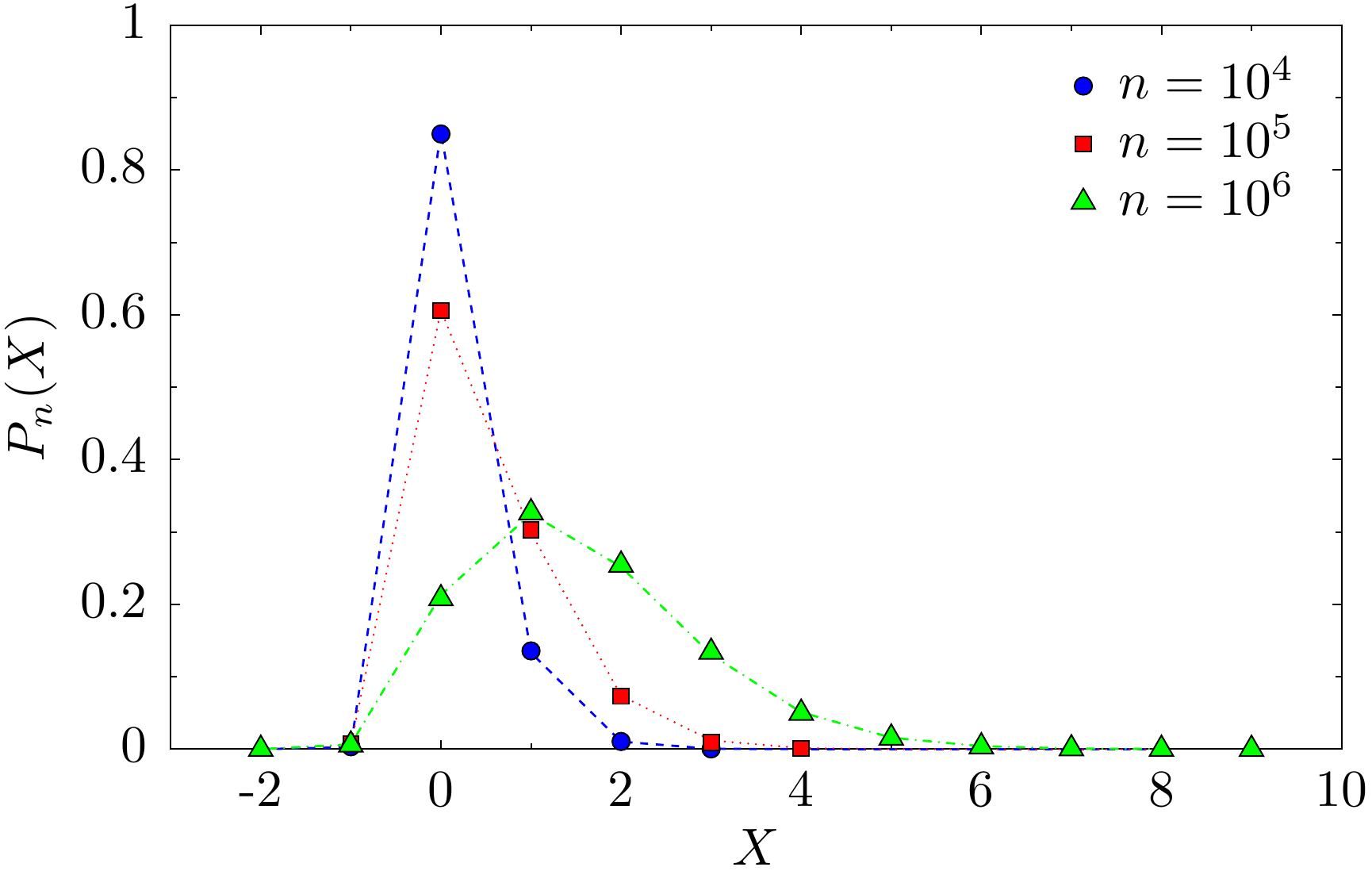}
\end{center}
\caption{The distribution $P_n(X)$  for $\rho_0 = 0.002$ and $p_1 = 0.98$. The dashed lines are our theoretical predictions in Eq. (\ref{diss}), while the symbols are the results of numerical simulations.}\label{fig3}
\end{figure}

Note finally that the regime of validity of our expressions with respect to the density $\rho_0$ is tested in  Fig. \ref{fig1}, where  we compare 
our theoretical predictions 
for the  
cumulants,  
obtained by the
 inversion of our general Eq. (\ref{gen2}), against
  the results of numerical simulations for different values of the 
  density $\rho_0$ of the vacancies, for different forces $f$ (defined as   $\beta f =\ln(p_1/p_{-1}$) and a fixed time moment $n = 100$. We observe a very good agreement 
 for very small values of $\rho_0$ and conclude that, in general, 
 the approach developed here provides a very accurate description of the TP dynamics for $\rho_0 \lesssim 0.1$. 
Further on, in Fig. \ref{fig2} we plot our theoretical predictions 
for the time-evolution of the cumulants for different values of the 
force and at a fixed density $\rho_0$. Again, we observe a perfect agreement between theory and simulations. Note that for small fields the reduced odd cumulants approach 1 from above, while for strong fields  from below. Last, we compare in Fig.\ref{fig3} our prediction in Eq. (\ref{diss}) against the numerical data and again observe a very good agreement between our analytical result and numerical simulations.

{\it Summary}. We have solved a minimal model of active transport in  crowded single-file environments. Our approach generalises the emblematic model of single file diffusion to the case of an active TP.  We have
derived explicit expressions, valid   in the limit of high density of bath particles,  of the full 
 distribution $P_n(X)$ of the TP position and of all its cumulants,  
 for arbitrary values of the bias $f$ and for any time $n$. Our analysis reveals striking features, such as the anomalous scaling $\propto\sqrt{n}$ of all cumulants, the equality of cumulants of same parity characteristic of a Skellam distribution and a convergence to a Gaussian distribution in spite of  asymmetric density profiles bath particles. Altogether, our results provide the full statistics of the TP position, and set the basis 
  for a refined analysis of real trajectories of active particles in  crowded single-file environments. 
 
 %We believe that these characteristics  provide a signature
 %of active transport in dense diffusing single file systems and could make possible its experimental identification and quantitative analysis.

 GO acknowledges fruitful discussions with Clemens Bechinger.  OB
is partially supported by the European Research Council
starting Grant FPTOpt-277998.  CMM and GO are partially supported by
the ESF Research Network "Exploring the Physics of Small Devices".

\end{document}